
\font\elevenmib   =cmmib10    \skewchar\elevenmib='177
\def\YUKAWAmark{\hbox{\elevenmib
 Yukawa\hskip0.05cm Institute\hskip0.05cm Kyoto \hfill}}

\def \frac#1#2{ { #1 \over #2} }
\magnification=1200
\baselineskip=6mm

\line{\hskip -6mm \YUKAWAmark \hfil YITP/K-996}
\vskip -2mm
\line{\hfil December 1992}

\vskip 1.5cm
\centerline{\bf MICROSCOPIC ORIGIN OF QUANTUM CHAOS IN ROTATIONAL
DAMPING}

\vskip 2cm

M. Matsuo$^{a)}$, T. D\o ssing $^{b)}$, E. Vigezzi$^{c)}$, and R.A.
Broglia$^{b,c,d)}$
\vskip 1cm
a) Yukawa Institute for Theoretical Physics, Kyoto University,  Kyoto 606-01,
Japan

b) The Niels Bohr Institute, University of Copenhagen,
Copenhagen, Denmark

c) INFN Sezione di Milano, Milano, Italy

d) Dipartimento di Fisica dell'Universit\`a di Milano, Milano, Italy

\vskip 2cm
\noindent {\bf Abstract}
\vskip 0.5cm
The rotational spectrum of $^{168}$Yb is calculated diagonalizing
different effective interactions within the basis of unperturbed
rotational bands provided by the cranked shell model. A transition
between order and chaos taking place in the energy region
between 1 and 2 MeV above the yrast line is observed,
associated with the onset of rotational damping. It can be related to
the higher multipole components of the force acting
among the unperturbed rotational bands.

\vfill
\eject

While the phenomenon of chaos is associated with a well defined
behaviour of classical systems which can be quantified mathematically
in terms of Poincar\'e maps and Ljapunov exponents, the proper
definition of quantal chaos is still a debated question. The standard
conjecture used in characterizing the phenomenon states that the
fluctuations of generic quantum systems, which in the classical
limit are fully chaotic, coincide with those of the gaussian orthogonal
ensemble (GOE) [1].
The words ``fluctuation properties" refer to
both level spacing and amplitudes.
As a rule, mean values of quantal many-body systems in general, and of
the
atomic nucleus in particular, do depend on the dynamics of the
system, that is, on the detailed properties of the Hamiltonian.
On the other hand, the fluctuations of the observables about their mean
values are governed by chaos and by the associated GOE stochasticity.
A basic question which still remains open is a proper identification
of the microscopic basis for this behaviour of many-body systems.

In the present paper we shall attempt at shading some light on this
issue, through the study of the damping of nuclear  rotational motion,
as
a function of the excitation energy (cf. Ref. [2] and refs. therein).
For this purpose we shall study the spectral and strength fluctuation
properties [3,4]  associated with the results obtained by diagonalizing
effective two-body  interactions in the basis set of unperturbed
rotational bands.
It will be concluded that the model in fact gives rise to  chaotic
behaviour, and that the high multipolarities of the interaction
are responsible for it.
Before proceeding further, we point out that in the following
we shall only consider the information about rotational
damping connected with single decays at moderate excitation energies
above the yrast. In other words, questions associated with
$\gamma - \gamma$ correlation and with motional narrowing [5,6]
will not be addressed.


The usual starting point of a microscopic description of the nuclear
spectrum is the shell model, where the nucleons move independently of
each other in an average potential. In fact, the mean field is the
organizing element of the variety of single-particle and collective
motion typical of the nuclear response.
Within this spirit, and assuming that the rotation
of the nucleus as a whole can be described in terms of
a classical external field, the cranked shell
model [7]
provides a convenient basis for a detailed description of nuclear
rotations as well as of rotational damping [5].
To be quantitative, a two-body residual interaction has to be
diagonalized in the basis of unperturbed rotational bands
as a function of the rotational frequency.
Most of the residual interaction which describes behaviour
which goes beyond the independent-particle model acts at the nuclear surface.
In other words, the nucleons move independently in the interior of the nucleus
and collide only when they are in the surface region. Of course,
there are also correlations in nuclear matter due to the short-range
repulsion in the nucleon-nucleon interaction. However, these
correlations
are nearly state-independent and can be neglected at low energies.
Among the possible
interactions the pairing and quadrupole model has been quite
successful in accounting for many systematic features of
nuclear levels [8].
If one is interested not only in the quadrupole but also in the octupole
and eventually hexadecapole response of the system, one should also
take into account the associated
multipole-multipole forces in the Hamiltonian.
It might seem that such an approximation is far too crude to be useful.
It turns out that in linear response theory, most of the interaction
between particles is irrelevant to the motion. The only important
part is that which produces an internal field with the same symmetry
as the external field exciting the motion. Thus, one writes the
interaction as a sum of multipole-multipole terms, and the only ones
of importance would be the terms with the correct multipolarity.

The simplest force which is both surface peaked and reproduces the
results of the multipole-multipole interaction is the surface-delta
interaction (SDI) [9]
$$
V(1,2) = \ 4 \pi V_o \delta(r_1 - R_o) \delta(r_2 - R_o)
\sum_{\lambda, \mu} Y_{\lambda, \mu}^* (\hat{r}_1)
Y_{\lambda, \mu} (\hat{r}_2)  \ \ . \eqno(1)
$$
The sum in (1) is, apart from a constant, a delta function in the angle
between $\vec {r}_1 $ and $\vec{r}_2$.
The coupling constant can be determined from self-consistent conditions
relating the density and the potential. These conditions are valid only
for those multipolarities producing waves with a wavelength larger than
the average distance between nucleons, that is,
$ \lambda \leq A^{1/3}$, where $A$ is the mass number.
For medium heavy nuclei one then obtains $\lambda \leq 5$.
This condition
is equivalent to the Debye frequency cut-off condition in the wavelength
associated with phonons in a lattice, and implies that the potential
and the density follow each other only for wavelengths at which the
many-body system reacts as a continuum medium.
For higher multipolarities (shorter wavelengths)
the motion of the particles will hardly give
rise to a dynamic mean field effect.

Consequently, one expects that the matrix
elements associated with the scattering of pairs of nucleons
through the low-multipole components of the force will be
dominant selectively  for the pairs with large spacial overlap.
On the other hand, the selection rules should be washed out
in collisions proceeding through the high-multipole components
of the interaction.
A confirmation of this expected behaviour is provided by Fig. 1,
where the off-diagonal matrix elements of a SDI and of a pairing plus
quadrupole force calculated in the cranked shell model basis
associated with $^{168}$Yb, are confronted. The
strong peak at $V =0$ for
the pairing plus quadrupole force
indicates the presence of
strong selection rules.

The associated matrices
have been diagonalized as a function of the rotational frequency
$\omega$, corresponding to  given  average angular momentum $I$.
The calculations were carried out for each parity and signature.
The unperturbed basis included the lowest
$10^3$ states. This truncation has been checked to give stable results
for the lowest 300 levels for each parity and signature. These cover
an interval in excitation energy of approximately 2.4 MeV above yrast.
The rotational strength or the stretched E2 transition strengths are
calculated between levels at the rotational frequency
$\omega$
and those at $\omega -2/\cal{J}$,
corresponding to $I-2$, where
$\cal{J}$ is the moment of inertia.
In the analysis of the strength fluctuations it is important
to measure them relative to a smooth strength function
averaged over many levels.
Consequently, we shall use in what follows relative E2 strengths
normalized in terms of the local mean value of the rotational
strength function (for details, cf. Ref.[10]).

The consequences of the differences observed between the SDI and the
pairing plus quadrupole matrix elements in the rigidity of the
spectrum as measured by the $\Delta_3$ statistics, as well as in the
normalized
strength distribution of stretched E2 transitions are very conspicuous,
as shown in Fig. 2.
For the sake of fair comparison, we used in this calculation a
pairing and quadrupole force whose strength (a factor 1.5 larger
than the one shown in Fig.1) gives the same
root-mean-square value of the off-diagonal matrix elements
as that of SDI.
Already at excitation energies of a couple of
MeV above the yrast line the SDI results for $\Delta_3$ and for the
$B$(E2) strength distribution resemble closely a Wigner surmise
and a Porter-Thomas distribution respectively.
A strong deviation from these limits is
found in the case of the pairing plus quadrupole model.
These results indicate that the high multipole
terms [11]
contained in the SDI have considerably less
selectivity than the lowest multipole terms common for
the SDI and for the pairing plus quadrupole interaction,
allowing the unperturbed rotational bands to interact
more democratically, irrespective of the participating
single-particle orbits involved.

For excitation energies below 1 MeV above the yrast line,
the $B$(E2) transitions essentially populate a single state, and one can
speak about a region of discrete spectroscopy.
Defining discrete band transitions as carrying more than 71 $\%$
of the total $B$(E2) strength, the model predicts
the presence of about 36 rotational bands consistent with
the number of 30 reported in the fluctuation analysis of the
rotational decay spectra [2].
Between 1 MeV and 2 MeV the rotational pattern breaks down
in the sense that the E2-decay out of a state of spin $I$
will not go to a unique state, but to a range of final states
at spin $I-2$. This behaviour is quite evident from
Fig.3, where
we display examples of the strength distribution associated
with stretched E2 transitions from selected states as a function
of the excitation  energy, as well as the branching number
$$
n_{branch} = \left (\sum_j  S_{ij}^2 \right )^{-1}
\eqno(2)
$$
where $S_{ij}$ is the fraction of the rotational strength of a level
$i$ at a spin $I$, going to a level $j$ at spin $I-2$.
This quantity gives an indication of
the number of states populated by the decay out of a single
state (when all the $S_{ij}$ are equal, it coincides with such number).
Branching number less than 2 marks the region of discrete bands, and
Fig.3 indicates a gradual transition between discrete and damped
rotation taking place around an energy region of
$U_0 \simeq 800$ keV above yrast. Presence of the discrete
rotational bands in the energy region near the yrast is reflected
also in the deviation in $\Delta_3$ from the GOE for the lowest
100 levels displayed in Fig.2.

We conclude that rotational damping and the associated GOE behavior
of the fluctuations of the system and of the transition strength
are intimately connected with the high multipoles present
in the residual interaction responsible for the mixing of the
unperturbed rotational bands.
While the study of the nuclear response has provided accurate
information concerning the low multipole components of the interaction,
very little is known about components with high
multipolarities.
In this sense, a detailed experimental
study of the statistical properties of the rotational spectrum
as a function of the rotational frequency may prove instrumental
to obtain a better understanding of this poorly known part of the
effective nuclear residual interaction.

\vskip 0.5cm
\noindent {\bf References}

\parindent =0pt

[1] O. Bohigas, M.J. Giannoni and C. Schmit, Phys. Rev. Lett.
{\bf 52}, 1 (1984).

[2] B. Herskind, A. Bracco, R.A. Broglia, T. D\o ssing, A. Ikeda,
S. Leoni, J. Lisle, M. Matsuo, and E. Vigezzi,
Phys. Rev. Lett. {\bf 68}, 3008 (1992);
B. Herskind, T. D\o ssing, S. Leoni, M. Matsuo, and E. Vigezzi,
Prog. Part. Nucl. Phys. {\bf 28}, 235 (1992), and refs. therein.

[3] Previous works have assumed from the
outset the system could be described in terms of GOE
wavefunctions [5] or that the unperturbed rotational
bands interact democratically with
constant matrix elements [4].

[4] S. \AA berg, Phys. Rev. Lett. {\bf 64}, 319 (1990);
S. \AA berg, Prog. Part. Nucl. Phys. {\bf 28}, 11 (1992).

[5] B. Lauritzen, T. D\o ssing and R.A. Broglia, Nucl. Phys.
{\bf A457}, 61 (1986).

[6] R.A. Broglia, T. D\o ssing, B. Lauritzen and B.R. Mottelson,
{\bf 58}, 326 (1987).

[7] R. Bengtsson and  S. Frauendorf, Nucl. Phys. {\bf A314}, 27 (1979);
{\bf A327}, 137 (1979).

[8] A. Bohr and B.R. Mottelson, {\it Nuclear Structure.} vol. II
(Benjamin, 1975).

[9] I.M.Green and S.A. Mozkowski, Phys. Rev. {\bf 139}, B790 (1965);
A. Faessler, Fort. der Physik {\bf 16}, 309 (1968).

[10] M. Matsuo, T. D\o ssing, B. Herskind and S. Frauendorf, in preparation.

[11] While there have been attempts at studying
states based on configurations with high spins through
particle transfer (cf. ref.[12]), these studies have
not provided much insight into the high multipole
components of the residual interaction. Note however the
studies carried out on high-spin states of nuclei like
$^{205}$Pb [13].

[12] P.D. Bond, J.Barrette, C.Backtash, C.E.Thorn, and A.J.Kreiner,
Phys. Rev. Lett. {\bf 46}, 1565 (1981);
P.D. Bond, in {\it Semiclassical Descriptions of Atomic and
Nuclear Collisions} (North Holland, 1985), p151.

[13] C.G. Lind\'en, I. Bergstr\"om, J. Blomqvist, K.-G. Rensfelt,
H. Sergolle, and, K. Westerberg, Z. Phys. {\bf A277}, 273 (1976);
J. Blomqvist, L. Rydstr\"om, R.J. Liotta, and C. Pomar,
Nucl. Phys. {\bf A423}, 253 (1984).

\vfill
\break
\noindent {\bf Figure caption}
\vskip 0.5cm

{\bf Fig.1} Distributions of the off-diagonal two-body matrix elements
of
the pairing plus quadrupole (P+QQ) force and the surface delta
interaction (SDI). The figure includes the two-body matrix elements
$V_{\alpha\beta\gamma\delta}$, where
$\alpha,\beta,\gamma,\delta$ denote
the cranked shell model single-particle orbits lying near the
Fermi surface within 3 MeV, defined at
$\omega=0.5$ MeV in $^{168}$Yb.
The strength of SDI is taken from  Ref.[9] ($F=27.5/A$ MeV)
while those of P+QQ are given by the selfconsistent values
with polarization effect (Ref.[10] for detail).
The dashed curve represents a
Gaussian distribution which gives the same root mean square
(18 keV) as SDI.

{\bf Fig.2} The ridigidy $\Delta_3$ of the energy levels
and  the distribution of the normalized rotational strengths
$s_{ij}=S_{ij}/\langle S_{ij}\rangle$.
The lowest 300 levels for each parity and
signature are grouped into three bins of 100 levels.
The $\Delta_3$ analysis is done seperately for these three bins,
which are marked by  $a$, $b$ and $c$
and whose excitation energies above yrast are approximately
0.0-1.9 MeV, 1.9-2.2 MeV and 2.2-2.4 MeV,
respectively for SDI, and 0.0-1.6, 1.6-1.9, and 1.9-2.1 MeV for P+QQ.
The dashed and the dash-dotted curves in the $\Delta_3$
plot correspond to the Poisson distribution and the GOE limit,respectively.
The normalized strength distribution is calculated for
the transitions from the $201 - 300$-th levels (belonging
to the third bin $c$) with
gamma-ray energies satisfying $0.90 < E_\gamma < 1.05$ MeV.
The dashed curve represents the Porter-Thomas
distribution.
The rotational frequency and the nucleus in this calculation
is the same as in Fig.1.

{\bf Fig.3} The branching number for decay out of
states is shown as a function of the excitation energy
above yrast, calculated with the surface delta interaction and
averaged within bins 100 keV wide. The
inset displays the distribution of rotational strength
from two typical levels of the low and high excitation energy.

\bye